\documentclass[twocolumn,apjl]{aastex631}
\submitjournal{ApJLett}
\usepackage{bm}
\shorttitle{Particle acceleration at oblique shocks}
\shortauthors{Kumar and Reville}

\begin{document}

\title{Non-thermal particle acceleration at highly oblique non-relativistic shocks}

\correspondingauthor{Naveen Kumar}
\email{naveen.kumar@mpi-hd.mpg.de}

\author[0000-0002-5132-7228]{Naveen Kumar}
\affiliation{Max-Planck-Institut f\"ur Kernphysik,
Saupfercheckweg 1, Heidelberg, 69117 , Germany}

\author[0000-0002-3778-1432]{Brian Reville}
\affiliation{Max-Planck-Institut f\"ur Kernphysik,
Saupfercheckweg 1, Heidelberg, 69117 , Germany}



\begin{abstract}

The non-thermal acceleration of electrons and ions at an oblique, non-relativistic shock is studied using large scale particle-in-cell (PIC) simulations in one spatial dimension. 
Physical parameters are selected to highlight
the role of electron pre-heating in injection and subsequent acceleration of particles at high Mach number shocks. 
Simulation results show evidence for the early onset of the diffusive shock acceleration process for both electrons and ions at a highly oblique subluminal shock.
Ion acceleration efficiencies of $\lesssim 5\%$ are measured at late times, though this is not the saturated value.
\end{abstract}



\section{Introduction} \label{sec:intro}

High Mach number, non-relativistic shocks are known to be strong sources of non-thermal radiation \cite[e.g.][]{Hinton:2009wo}. Multi-wavelength measurements are thought to be a signature of diffusive shock acceleration (DSA) \cite[see][for reviews]{Blandford:1987uy,Bell:2014tw}. Crucially, DSA requires frequent scattering on electric/magnetic field fluctuations to maintain near isotropy of the particle distribution. If particles interacting with the shock can escape the thermal pool in sufficient quantity, the mechanism is self-sustaining \citep{Eichler}. The plasma processes that initiate injection into DSA are not yet fully understood, though it is clear the ambient conditions, in particular the magnetic field obliquity, play an important role \cite[e.g.][]{Kirk:1989wb,Caprioli:2014tg, Guo:2015tt}. If the shock propagation and magnetic field directions are uncorrelated, it follows that the mean obliquity, i.e. the angle between the upstream magnetic field and shock normal direction, is $\langle{\theta}_{\rm n}\rangle ={60}^{\circ}$. For a given shock velocity $\upsilon_{\rm sh}$, particles can not outrun the shock upstream along fieldlines if $|\cos ~\theta_{\rm n}|< \upsilon_{\rm sh}/c$ (so-called superluminal shocks). However,  
for non-relativistic shock velocities this transition occurs at large angles: $\theta_{\rm n}\lesssim 90^\circ$. Thus characterising the ability of shocks with $\theta_{\rm n} > {60}^{\circ}$, (approximately half of all shock surfaces) to inject electrons and ions from the thermal pool and subsequently accelerate to high energies is of interest to the high-energy astrophysics community.

That highly oblique shocks can accelerate particles via DSA is not controversial. $\gamma$-ray observations of the binary stellar system $\eta$-Carinae provide compelling evidence that both ion and electron acceleration are ongoing at the oblique shocks produced in the wind collision region \citep{White:2020we}. The relative contribution of electrons and ions to the $\gamma$-ray emission from supernova remnants is unclear. However, the X-ray synchrotron rims that outline almost the entire circumference of young remnants such as Cassiopeia A \citep{Patnaude:2014tr} and Tycho \citep{Eriksen:2011wg} indicate not only that electrons are accelerated locally to $\gg$ TeV energies, but that acceleration occurs over a large range of obliquities. A counter-example is SN 1006 for which non-thermal emission is more prominent in the quasi-parallel regions (the \lq\lq polar caps\rq\rq) of the supernova remnant \citep{Reynoso:2013tr}. However, the reasons underlying this asymmetry are not conclusive \citep{Winner20}, and may result from a combination of effects related to the ambient plasma conditions, the acceleration process itself or an inability to amplify magnetic fields sufficiently, and not necessarily a consequence of injection efficiency \cite[see for example][]{Bell:2011wn,Reville:2013vs}.

The physical mechanisms that determine the injection of electrons and ions into the acceleration process at magnetised shocks are distinct. The large difference in their respective masses imply that the relevant waves-particle resonances fall in different frequency ranges.
In particular, for electrons to participate in DSA, a pre-heating mechanism to energies $\sim m_{\rm i} u_{\rm sh}^2/2$, is required to facilitate resonance with low frequency MHD waves.  
This problem is often referred to as the ``injection problem". 
Various mechanisms such as electrostatic Buneman instability, oblique whistler waves or Weibel instability, have been proposed as potential solutions~\citep{Amano:2009aa,Muschietti:2017wv,Amano:2010uu,Riquelme:2011aa,Lembege:1990tb,Oka:2017aa,Bohdan:2019aa,Bohdan:2019ab}.

Kinetic plasma simulations, using particle-in-cell (PIC) or hybrid codes ~\cite[e.g.][]{Riquelme:2011aa,Caprioli:2014tg,Guo:2014aa,Kato:2015aa,Park:2015aa,Bohdan:2019aa,Bohdan:2019ab,Fang:2019aa,Hanusch:2020aa,Xu:2020aa}, can elucidate the physical processes underpinning the energy dissipation, and related non-thermal phenomena.
While non-thermal particle acceleration is found to occur in many high Mach number shock simulations, previous numerical studies of highly oblique shocks ($\theta_{\rm n}>60^\circ$) despite exhibiting efficient electron acceleration,
have yet to reveal significant non-thermal ion acceleration
\citep{Riquelme:2011aa,Caprioli:2014tg,Xu:2020aa}. 

In this Letter we explore the role of electron and ion kinetic scales separation in a realistic system, while capturing a sufficiently large number of ion-gyration times to demonstrate the onset of non-thermal ion acceleration. The large number of particles-per-cell (ppc) needed to counter the energy loss caused by plasma drag forces~\citep{Kato:2013uk,May:2014wi} restricts the current study to 1D3V (one dimension in space and three in momentum) simulations. In the  following we address the aforementioned issues and study both electron and ion acceleration at an oblique shock. The numerical set-up is summarised in the next section. Results are presented in section \ref{sec:results}, while our main conclusions are discussed in section \ref{sec:conclusions}.

\section{PIC Simulation setup}

For the simulations presented, we use the open-source PIC code SMILEI~\citep{derouillat2018smilei}. We employ a relaxation method of shock excitation by initializing the simulation in the shock-normal frame (SNF). The Rankine-Hugoniot (RH) jump conditions determine the downstream parameters (superscript ${\rm d}$) for our desired upstream conditions (superscript ${\rm u}$) ~\citep{Tidman:1971uq,Leroy:1981aa,Umeda:2014wj}. Unless stated otherwise, time and length are normalized to the inverse of the electron plasma frequency ($\omega_{\rm pe}^u=\sqrt{4\pi n_{\rm e}^u e^2/m_{\rm e}}$) and skin-depth ($c/\omega_{\rm pe}^u$), where $n_{\rm e}^u=1$ is the upstream electron plasma density, $c$ is velocity of light in vacuum, $m_{\rm e}$ $(-e)$ is the electron mass (charge). Velocity, momentum, temperatures and fields are normalized to $c$, $m_e c$, $m_{\rm e} c^2$ and $m_{\rm e}\,\omega_{\rm pe}^u c/e$, respectively. 

Making use of the following relations for the plasma beta (ratio of thermal to magnetic pressure) $\beta_{\rm e}^{\rm u}=4\pi n_{\rm e} k_{\rm B} T_{\rm e}/(B^ {\rm u})^2= 2(M_{\rm A}/M_{\rm S})^2$, electron cyclotron to plasma frequency ratio $\omega_{\rm ce}^u/\omega_{\rm pe}^u=\upsilon^{\rm u}(m_{\rm i }/m_{\rm e})^{1/2}/M_{\rm A}$, and electron thermal velocity $\upsilon_{\rm th,e}^u=(M_{\rm A}/M_{\rm S})(\omega_{\rm ce}^u/\omega_{\rm pe}^{\rm u})$, the upstream and downstream parameters are easily expressed in dimensionless unit, again using the RH jump conditions. Here $\upsilon_{\rm A}=B^{\rm u}/\sqrt{4\pi n_0 m_{\rm i}}$ the Alfv\'en velocity, while $M_{\rm A}=\upsilon^{\rm u}/\upsilon_{\rm A} =63$ and $M_{\rm S}=\upsilon^{\rm u}/\sqrt{(T_{\rm e}^{\rm u}+T_{\rm i}^{\rm u})/m_{\rm i}}=70$ are the Alfv\'enic and sonic Mach numbers, respectively. The ion mass is denoted by $m_{\rm i}$, and $T_{\rm e}^{\rm u},\, T_{\rm i}^{\rm u}$ are the electron and ion temperatures. The drift velocity of upstream plasma (or equivalently the shock velocity) is $\upsilon^{\rm u}=0.15~\hat{\bm{x}}$. Furthermore, we assume  $\beta_{\rm e}^{\rm u}=\beta_{\rm i}^{\rm u} \approx 1.8$, implying ion to electron temperature ratio $T_{\rm i}^{\rm u}/T_e^{\rm u}=1$, and $T_{\rm i}^{\rm d}/T_{\rm e}^{\rm d}=4$ in accordance with previous studies on collisionless shocks~\citep{Umeda:2014wj,Leroy:1981aa}. Different downstream temperature ratios were explored and early time results were comparable. Previous numerical shock studies ~\citep{Leroy:1981aa,Park:2015aa,Kato:2015aa,Fang:2019aa,Hanusch:2020aa} have focused on the regimes of high and low plasma betas. The resulting wave analyses in these regimes are more amenable to theoretical investigation. Here we focus on the intermediate regime $\beta \sim 1$, since extreme limits are in general specific to given astrophysical environments, which is beyond the present scope. The generation of plasma waves is sensitive to $\beta$, and consequently the electron heating mechanisms are expected to show a complex character in the intermediate plasma beta regime. We choose the upstream magnetic field, $\mathbf{B}^{\rm u}=B_{\rm x}^{\rm u}\hat{\bm{x}}+B_{\rm z}^{\rm u}\hat{\bm{z}}$ making an angle of $\theta_{\rm n} = 75^{\circ}$ with the shock normal.  We assume a reduced ion to electron mass ratio $m_{\rm i}/m_{\rm e}=256$, and grid and time resolutions of $\Delta x=0.1$ and $\Delta t = 0.095$, respectively. 
At $t=0$, the shock discontinuity is initialized with a thickness (in normalized units) $\delta x = 2\rho_{\rm i}= 2(\upsilon^{\rm u}/B^{\rm d}) (m_{\rm i}/m_{\rm e})$ with hyperbolic tangent profile. The total length of the simulation domain is $L_{\rm x} \approx 570 \rho_{\rm i}$.  \textcolor{black}{As mentioned previously, plasma drag can inhibit electron energization if ppc number is insufficient \citep{Kato:2013uk, May:2014wi}. Using Eq.(37) of \cite{Kato:2013uk}, we conservatively estimate a minimum requirement for ppc to be $\gg 100$. Consequently, we use $512 (1024)$ ppc per species for the upstream (downstream) population. The global energy conservation was satisfied to better that $1\%$ throughout the simulation.}

\section{Results}
\label{sec:results}

\begin{figure}
\centering
\includegraphics[width=1.0\linewidth]{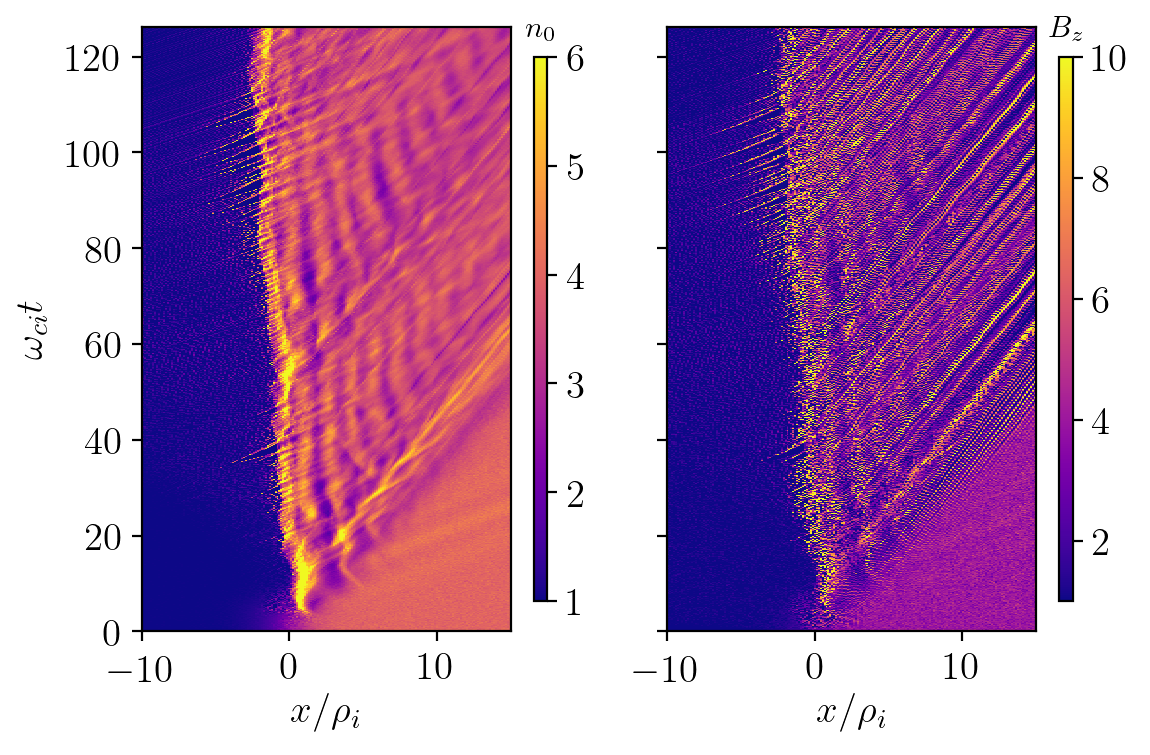}
\caption{Zoomed-in view of the ion density map and the  transverse magnetic field. Time and the length are normalized to ion gyro frequency ($\omega_{\rm ci}=(m_{\rm e}/m_{\rm i})B^{\rm d}$) and ion gyro radius ($\rho_{\rm i}$) respectively. The ion number density is normalised to its initial upstream value, and similarly the magnetic field is normalized to the initial upstream value, $B^{\rm u}_z$.}
\label{fig:maps}
\end{figure}

Figure \ref{fig:maps} shows the spatio-temporal evolution of the plasma density and transverse ($B_z$) magnetic field. It can be seen that the initial magneto-hydrodynamic (MHD) discontinuity evolves into a well-defined shock structure, separating the upstream and downstream plasmas, and the creation of a shock overshoot region. The transverse magnetic field follows a similar pattern to the plasma density. Periodic magnetic field reformation occurs at the shock at approximately the ion gyro-frequency (corresponding to the downstream magnetic field)~\citep{Umeda:2014wj,Leroy:1981aa,Lyu:1990ui,Lembege:1990tb,Burgess:2015aa}. \textcolor{black}{Magnetic field reformation is generally associated to insufficient dissipation in the shock ramp region leading to ion reflection. In the case of quasi-parallel shocks for example, this results in the generation of high-frequency whistler waves~\citep{Lyu:1990ui}.} The transverse magnetic field seen in Fig.\ref{fig:maps} shows strong filamentary structure in the downstream region. In both panels, one may also notice (for $\omega_{\rm ci}t > 30$), the appearance of sharp features extending in the upstream region. 
These structures are generated in the shock foot region, excited by the reflection of incoming ions, which are however both electromagnetic and electrostatic in nature, \textcolor{black}{in contrast to expectations of quasi-parallel shocks \citep{Lyu:1990ui}.}
These waves play an important role in the energization of the electrons as we show later in Fig.\ref{fig:testelectrons}.

\begin{figure}
\begin{center}
\includegraphics[width=0.95\linewidth]{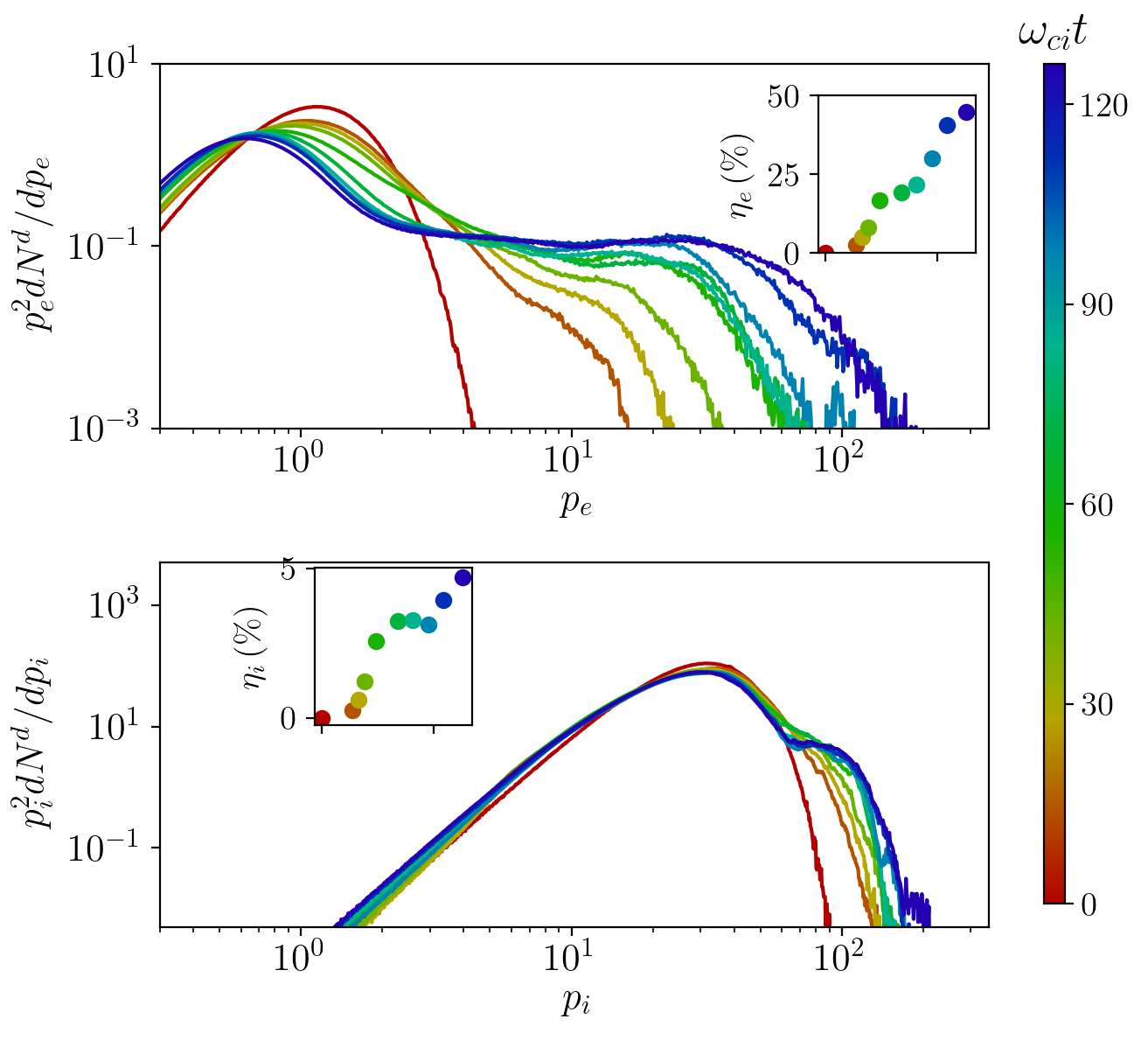}
\caption{Downstream spectra of electrons (upper panel) and ions (lower panel) at different times in the simulation. Both $p_{\rm e}$ and $p_{\rm i}$ are normalized to $m_e c$. The insets show the evolution of energy in the non-thermal populations of the respective particle species with time. Note $\eta_{\rm e,i}$ is the ratio of integrated kinetic energy in the non-thermal particles relative to thermal particles.}
\label{fig:spectra}
\end{center}
\end{figure}

Fig.\ref{fig:spectra} shows the evolution of the momenta distribution for both species in the downstream. The spectra is summed from $x_{\rm i} \approx 2\rho_{\rm i}$ to $x_{\rm f} \approx 16 \rho_{\rm i} $ in the downstream direction and have been multiplied by $p^2$ to highlight the emergence of flat non-thermal tails at high energies. At the end of the simulation, an appreciable fraction of electrons have acquired sufficient momentum for their gyro radii to exceed the shock thickness (in the range $p_{\rm e} \sim 10 {-} 40$). These electrons can participate in the DSA process. Once this threshold is reached (at $\omega_{\rm ci}t\approx 50$) the emergence of a distinct population above this momentum is visible, although this appears to be a transient effect. The growth of a non-thermal ion population is also evident, although the separation from the thermal population is not as developed as that of the electron population, due to the limited number of gyro-periods captured.
The onset of the non-thermal ion population develops in earnest only after the non-thermal electrons have acquired gyro-radii comparable to that of the bulk of the ions, emphasising the necessity for long run times. This is also indicated by features in the time evolution of the acceleration efficiency $\eta_{\rm e,i}$,
defined here as the ratio of the kinetic energy in the thermal and non-thermal components.
For convenience we define the non-thermal component to be the integrated kinetic energy density of particles with $p_{\rm e}>4$ in the case of electrons and $p_{\rm i} >90$ for ions (see Figure \ref{fig:spectra}).
At the end of the simulation, the acceleration efficiency of electrons, according to our definition, is $\eta_{\rm e} \approx 45\%$ while for the ions it is $\eta_{\rm i} \approx 5\%$ (see insets). 
The large value for the non-thermal electron energy efficiency is the combined effect of increasing maximum energy, and depletion of the thermal component. Relative to the incoming upstream kinetic energy density, this efficiency would remain small. The ion energy efficiency on the other hand is non-negligible, and crucially shows no sign of saturating at the end of the simulation.  



\begin{figure}
\begin{center}
\includegraphics[width=1.05\linewidth]{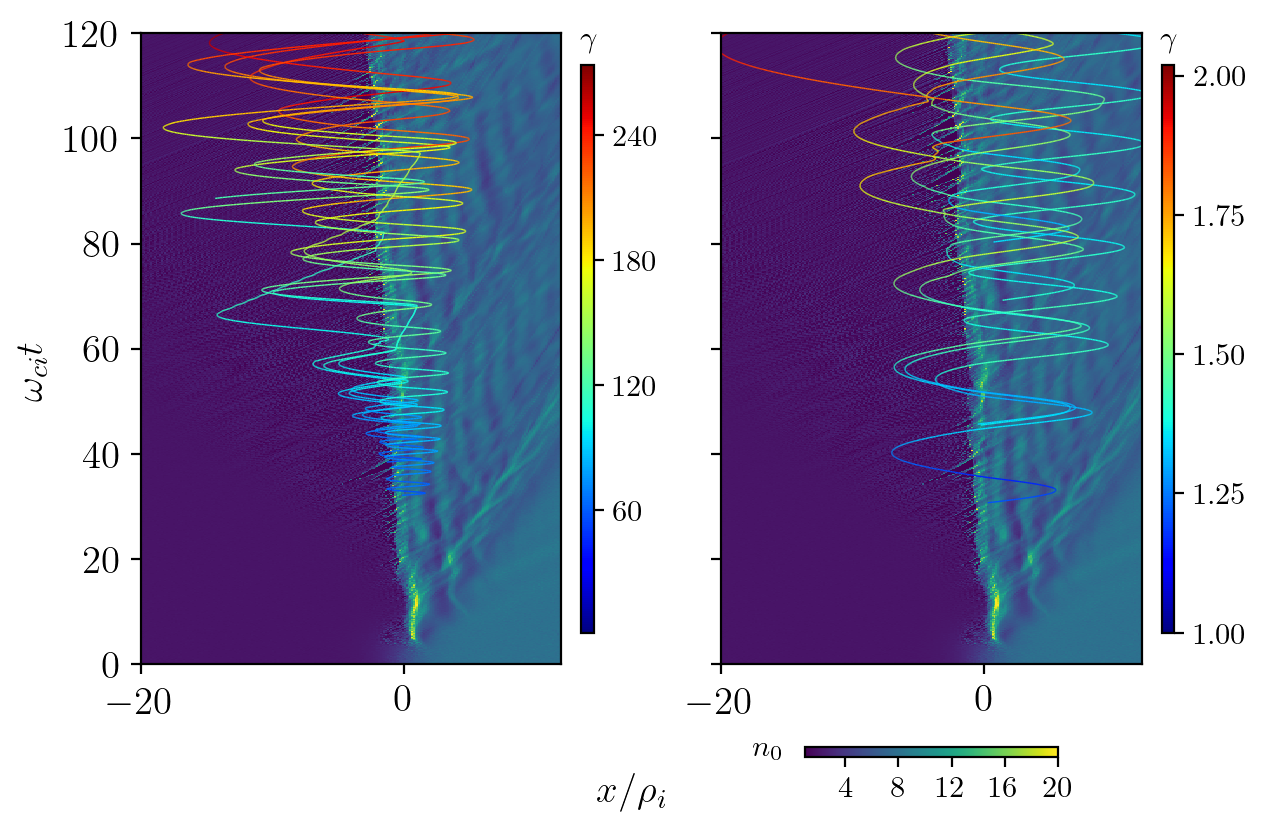}
\caption{Zoomed-in view of the trajectories of five energetic electrons (left) and ions (right). The horizontal colorbar shows the plasma density maps associated with the shock, while the vertical colorbars show the Lorentz factor $\gamma$ of respective species. Here $n_{\rm 0}=n_{\rm e} + n_{\rm i}$ represents the cumulative densities of all populations. }
\label{fig:ei_trajec}
\end{center}
\end{figure}

\begin{figure}
\begin{center}
\includegraphics[width=1.0\linewidth]{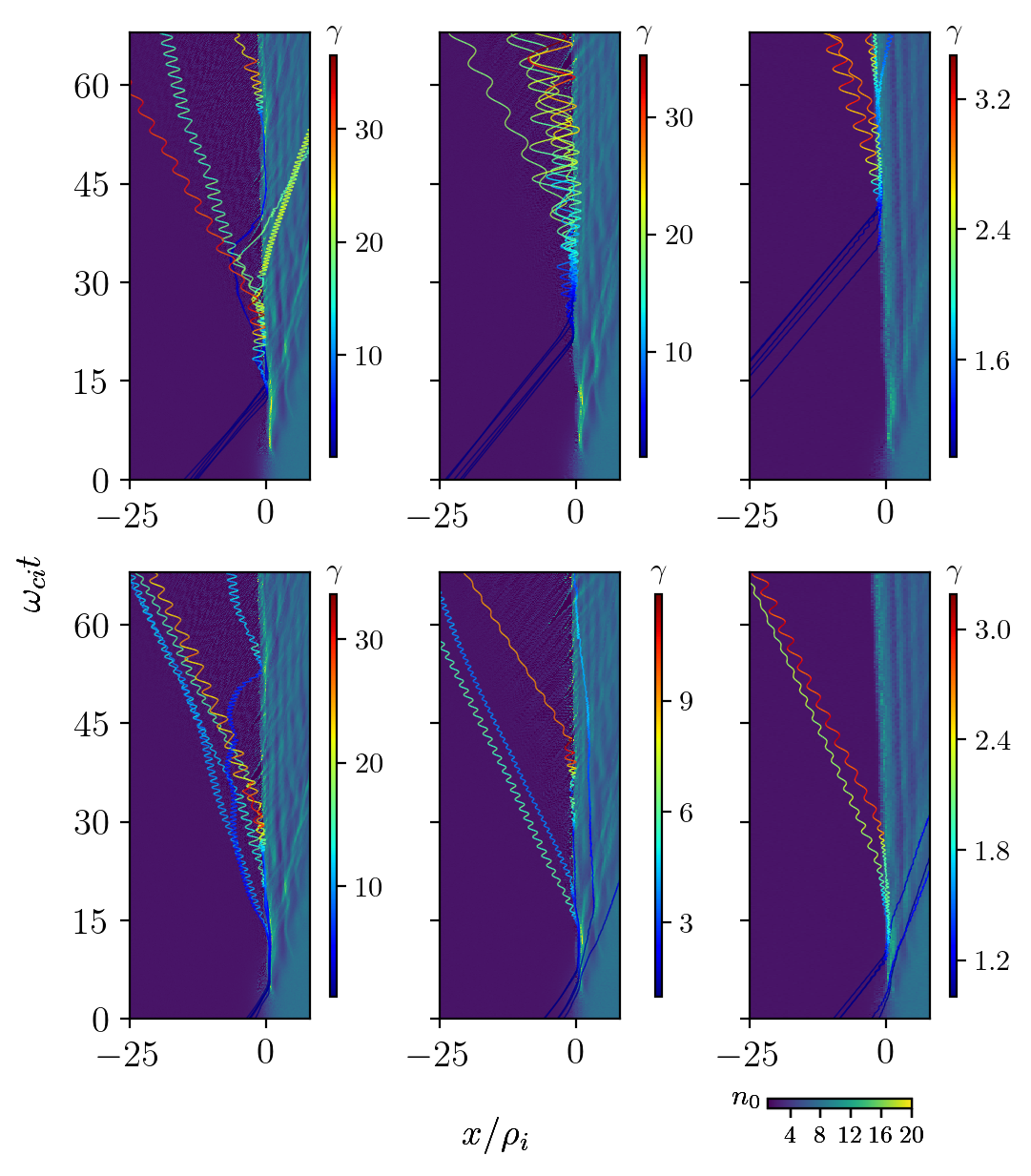}
\caption{Trajectories of five most energetic test electrons for different mass ratios. First, second and third columns are for $m_{\rm i}/m_{\rm e}=256, 100, 25$, respectively. Upper and lower rows show two populations of test electrons initialized in the upstream plasma further (upper row) and closer (lower row) to the shock discontinuity in the simulation box. Here $n_0$ is same as in Fig.\ref{fig:ei_trajec}.}
\label{fig:testelectrons}
\end{center}
\end{figure}

Fig.\ref{fig:ei_trajec} shows the trajectories of the five most energetic electrons and ions from the simulation corresponding to Fig.\ref{fig:spectra}. Particles are tracked only after crossing a population specific threshold condition. 
As noted earlier, the ion acceleration is negligible prior to electrons acquiring sufficient energies to have their gyro-radii comparable to that of the drifting ions. A causal connection between the two however, has not been established. To reach these energies, a continuous energization process (or processes) is required, which we address below (see Fig.\ref{fig:testelectrons}). Such pre-heating/energization mechanisms are unnecessary for ions, since their gyro-radii are already of the order of the shock thickness. The highest energy ions are seen to make large excursions into the upstream direction and display features of DSA acceleration. The guiding centre for example undergoes diffusive like behaviour in the turbulent pre- and post-shock regions. Note that by selecting the five highest energy particles, we are biased towards particles that have remained close to the shock, and therefore undergone the largest number of acceleration cycles. At the end of the simulation, many energetic particles penetrate more than a gyro-radius upstream. Particles that venture far upstream have large negative $z$- component of their velocity, as required to escape along the field-lines.

To explore the electron energization, including the effects of ion to electron mass ratio, we tracked the evolution of two populations of test electrons in simulations with different mass ratios. \textcolor{black}{These test electrons are always present in the full simulation run and evolve according to the self-consistent time-dependent electromagnetic fields. However, they do not contribute to the charge and current densities used to update these fields.}
Each population of test electrons contains 1000 particles, and are initialized with the same plasma parameters as upstream electrons \textcolor{black}{viz. charge, mass, temperature, and velocity.} Each group is distributed in a span of $\delta x =1000$ in the upstream region, at a fixed distance from the shock. Fig.\ref{fig:testelectrons} reveals the dependence of electron energization ($\gamma$ factor) on the mass ratio, a consequence of the larger cross-shock potential. One may also note a few additional trends: first the population of electrons that arrive later in time (upper row) generally display a larger energy gain compared to the population that arrives earlier (lower row). Energization is due to both the cross-shock potential and the interaction with the high frequency electrostatic and electromagnetic waves traveling upstream, as has been previously discussed~\citep{Leroy:1981aa, Ligorini:2021wx,Umeda:2014wj,Muschietti:2017wv}.  Since the Alfv\'en Mach number in our simulations satisfies the condition $M_{\rm A} \geq (\textrm{cos}\,\theta/2) \sqrt{(m_{\rm i}/m_{\rm e})\beta_{\rm e}^{\rm u} }$, generation of Buneman waves and oblique whistler modes are expected~\citep{Amano:2010uu}. These waves have ample time to grow in magnitude and transfer energy to the later population of electrons (upper row). Since the first group of test electrons interact with these waves earlier in the process, they do not gain as much energy. The second trend is that, superimposed on the gyromotion of the electrons, pitch-angle scattering features are observable. The effect is most prominent in the large mass-ratio simulations at later times (upper-left most panel) indicating that, similar to the cross-shock potential, the strength of the waves is dependent on the mass-ratio.

\section{Conclusion and Discussion}
\label{sec:conclusions}

We have investigated oblique, subluminal non-relativistic shocks using PIC simulations, and found they are capable of accelerating both electrons and ions. At the end of our simulation, a moderate energy-conversion efficiency $\sim 5\%$ is found for the ions, although this is not its saturated value. Early-time results from the simulation are broadly consistent with those found in previous studies in a comparable regime \citep[e.g.][]{Xu:2020aa}, \textcolor{black}{using a different shock initialization approach (piston-driven shock in the upstream rest-frame)}. 
Several previous works find that electron acceleration is efficient for  oblique high-Mach-number shocks, though the rate of acceleration decreases above $p_{\rm e} \approx u_{\rm sh} m_{\rm i}/m_{\rm e}$. We verify this trend (see Fig.\ref{fig:ei_trajec}) that electrons are \lq\lq rapidly\rq\rq\, energized to the same momentum, at which point they can interact with ion-scale fluctuations. However, as the gyration periods (a proxy for the acceleration time) above this momentum are now comparable, longer simulation times reveal that the spectral cut-offs march approximately in tandem, a feature not previously discussed. At the end of the simulation, the ions are trans-relativistic, and thus their increased inertia (and gyration period) result in less frequent shock-crossings. Acceleration to energies beyond what is achieved here thus require significantly longer run-times, which is not currently feasible. This presents a challenge to determining the saturated ion acceleration efficiency.
While the ion efficiencies presented are still low compared to predictions from quasi-parallel shocks configurations \cite[e.g.][]{Caprioli:2014tg,Park:2015aa}, \textcolor{black}{the most energetic particles continue to penetrate upstream. In this case, we expect the accelerated-particle mediated precursor to continue growing in extent and amplitude with time.} 
If this holds, we speculate that the conditions close to the viscous shock (i.e. the position of the MHD discontinuity) become increasingly turbulent, with particle transport in the shock vicinity being largely independent of the initial field orientation \cite[cf.][]{Reville:2013vs}. 

Identification of a causal connection between ion and electron acceleration has not been established. In our simulations, a transition in the acceleration efficiency appears soon after the maximum energy electrons have comparable gyro-radius to the incoming ions. This indicates that the acceleration of the two species are intimately linked. Whether further electron acceleration bootstraps ion acceleration, or vice-versa, is unclear. 
The initial energization of electrons from the thermal pool appears to be sensitive to the phase at which they first interact with the (reforming) shock. Favourable electrons are reflected and can subsequently interact with the high-frequency waves in the shock-foot region, where they undergo further heating. The stopping and reflection of the ions at the shock ramp is responsible for driving these high frequency electrostatic and electromagnetic perturbations in the upstream incoming plasma~\citep{Amano:2010uu}. 
We confirm that the interactions of test electrons with these waves in the immediate shock upstream is the primary channel for heating. 
Noticeable differences are visible between test electrons that interact early in the shock evolution, compared with those that arrive later, when these waves are further developed. It is not possible however to attribute the initial electron energization to one single instability or a single physical process for our simulation parameters. 

The generation of waves in the upstream region depends not only on the Alfv\'en/sonic Mach numbers, and mass ratio, but at early times appears also to be sensitive to the initial setup of the numerical simulations, which we discuss below~\citep{Amano:2010uu,Amano:2012vi,Bohdan:2017aa,Muschietti:2017wv,Umeda:2014wj,Oka:2017aa,Riquelme:2011aa,Xu:2020aa,Ligorini:2021wx}. 
\textcolor{black}{
The threshold Alfv\'en Mach number for triggering these waves plays an important role in astrophysics, particularly in systems with variable shock conditions. \citet{White:2020we} recently interpreted the quenching of electron heating close to periastron passage of the binary system $\eta$-Carinae to occur when the shock Mach number falls below the critical whistler Mach number $M_{\rm A,crit}=\sqrt{m_{\rm i}/m_{\rm e}}$, although different predictions can be found in the literature; e.g. \citet{Amano:2010uu} suggest $M_{\rm A} = (\textrm{cos}\,\theta/2) \sqrt{(m_{\rm i}/m_{\rm e})\beta_{\rm e}^{\rm u} }$, while conversely others find whistler waves favour low Mach shocks $M_{\rm A} \lesssim \sqrt{m_{\rm i}/m_{\rm e}}$ \cite[see the 2D simulations of][and references therein]{Riquelme:2011aa}. Our current simulations exceed both thresholds, although a parameter survey similar to that of \citet{Xu:2020aa} is needed. Irrespectively, our results have important consequences for the interpretation of astrophysical observations.}

\textcolor{black}{In this work we focus on simulations in the SNF 
\cite[see][for a summary of different shock initialization methods]{Burgess:2015aa,Lembege:1990tb}.
Whether simulation results specific to non-thermal particle acceleration are sensitive to the shock launching mechanism is an open question, as the initial development is unavoidably sensitive to this choice. Previous simulations in the downstream frame, where the incoming plasma is reflected off a wall, the shock formation is mediated by different beam-plasma instabilities~\citep{Riquelme:2011aa,Kato:2015aa}. Nevertheless, our intermediate time results bear striking similarity with those of \citet{Xu:2020aa}, who perform simulations in the upstream rest-frame. 
This suggests the long-term evolution of the shock dynamics and particle acceleration is the same in different frames using alternative setups.}
A more direct comparison can be made between the present results, performed in the SNF, with matching simulations (using the relaxation setup) performed in the de-Hoffman and Teller frame (dHT) \citep{KumarICRC}. This frame is ideal for acceleration studies since both the shock \emph{and} the zeroth-order field lines are static (in the SNF, field lines are continuously dragged across the shock). These two frames are related by an appropriate Lorentz boost along the shock surface. While the runtimes for results presented in \citet{KumarICRC} were shorter, compatible spectra are found at these times, when corrected for frame transformations.\footnote{These results are to be presented in an upcoming long paper.} However, 
in Lorentz transforming, the 1D simulation probes a different $\bm{k}$-vector of the upstream plasma dispersion. This motivates the necessity for simulations in more than one dimension \cite[cf.][]{Riquelme:2011aa}. Thus, further large-scale higher dimensional simulations are required to shed more light on the results presented in this Letter.

\begin{acknowledgments}
We thank J. Kirk for many stimulating discussions. 
We gratefully acknowledge the use of Raven and Cobra supercomputers of Max-Planck Central Data Facility in Munich. We thank the technical staff especially Markus Rampp for the support.
\end{acknowledgments}


\bibliographystyle{aasjournal}



\end{document}